\documentclass[a4paper, 10pt, conference]{IEEEtran}
\usepackage{blindtext, graphicx}
\usepackage{amssymb}

\usepackage{multirow}
\usepackage{graphicx}
\usepackage{algpseudocode}
\usepackage{cite}
\usepackage{amsmath}

\usepackage[ruled,vlined,commentsnumbered,linesnumbered,lined,boxed]{algorithm2e}

\usepackage{tikz}
\usepackage{siunitx}

\usepackage{tabularx}

\begin{document}

%
\title{\vspace{0.0in} CoV-TI-Net: Transferred Initialization with Modified End Layer for COVID-19 Diagnosis}
\author{\IEEEauthorblockN{Sadia Khanam$^a$, Mohammad Reza Chalak Qazani$^b$, Subrota Kumar Mondal$^{c, d}$, H M Dipu Kabir$^b$, \\  Abadhan S. Sabyasachi$^{c, e}$,  Houshyar Asadi$^b$, Keshav Kumar$^f$, Farzin Tabarsinezhad$^g$,\\ Shady Mohamed$^b$, Abbas Khorsavi$^b$, Saeid Nahavandi$^{b, h}$}\\ \IEEEauthorblockA{ $^a$Dhaka Dental College, Dhaka, Bangladesh. $^b$Institute for Intelligent Systems Research and Innovation (IISRI), \\Deakin University, Australia. $^{c}$ The Hong Kong University of Science and Technology, Hong Kong. \\$^d$Macau University of Science and Technology, Macao.
$^e$Manipal University Jaipur, Rajasthan, India.
\\$^f$University Institute of Computing, Chandigarh University, Punjab, India.
$^g$University of Tehran, Iran.
\\$^h$Harvard Paulson School of Engineering and Applied Sciences, Harvard University, Allston, MA 02134 USA.\\\{m.chalakqazani, hussain.kabir, shady.mohamed, abbas.khosravi, saeid.nahavandi\}@deakin.edu.au} }

\maketitle

\begin{abstract}
This paper proposes transferred initialization with modified fully connected layers for COVID-19 diagnosis. Convolutional neural networks (CNN) achieved a remarkable result in image classification. However, training a high-performing model is a very complicated and time-consuming process because of the complexity of image recognition applications. On the other hand, transfer learning is a relatively new learning method that has been employed in many sectors to achieve good performance with fewer computations. In this research, the PyTorch pre-trained models (VGG19\_bn and WideResNet -101) are applied in the MNIST dataset for the first time as initialization and with modified fully connected layers. The employed PyTorch pre-trained models were previously trained in ImageNet. The proposed model is developed and verified in the Kaggle notebook, and it reached the outstanding accuracy of 99.77\% without taking a huge computational time during the training process of the network. We also applied the same methodology to the SIIM-FISABIO-RSNA COVID-19 Detection dataset and achieved 80.01\% accuracy. In contrast, the previous methods need a huge compactional time during the training process to reach a high-performing model. Codes are available at the following link: github.com/dipuk0506/SpinalNet
\end{abstract}


%
\IEEEpeerreviewmaketitle

\section{Introduction}
The classification and categorization of the complex data including text, image, video, and document are great challenges in science-related sectors because it depends on various components \cite{wang2012end, jospin2022hands, qazani2020prepositioning}. Deep learning architecture such as deep neural network (DNN), conventional neural network (CNN), and recurrent neural network (RNN) are recently applied in classification to address this challenging issue \cite{chung2014empirical, qazani2019model, lee2009convolutional}.
The supervised learning-based model is used for classification to practically label the data. Naïve Bayes Classifier (NBC) is a simple technique in a supervised learning-based classification problem that is proposed by Rish \cite{rish2001empirical} and Murphy \cite{murphy2006naive}. It has been used in information retrieval and text recognition applications. NBC normally employs a bunch of words for feature extraction, then the order of the sequence cannot be reported on the results. Support Vector Machines (SVM) \cite{wang2005support, pedrammehr2018novel} is a quite popular classifier that has been used in dealing with a wide range of data. Yu and Joachims \cite{yu2009learning} proposed a new SVM structure using a latent variable to increase the performance of the model in three main sectors, including coreference resolution, motif-finding, and optimization. Tong et al. \cite{tong2001support} introduce a new approach by combining SVM and active learning. SVM has an extreme computational load during the training process that reduces the applicability of the model. Kabir et al. \cite{kabir2015bangla} proposed the Stochastic Gradient Descent (SGD) classifier to lease the computational burden of the system.
CireAan et al. \cite{ciregan2012multi} proposed the multi-column DNN for classification purposes. Jindal et al. \cite{jindal2019effective} proposed an efficient text classification model via a nonlinear processing layer during the training process. Krizhevsky et al. \cite{krizhevsky2012imagenet} employed 2-dimensional conventional layers with the addition of 2-dimensional space features of the image for classification purposes. Also, Hassan \cite{hassan2017deep} used CNN in text classification with the ability to reach convincing accuracy. RNN is employed in language processing and document classification applications by Mikolov \cite{mikolov2010recurrent} and Yang et al. \cite{yang2016hierarchical}, respectively. SpinalNet architecture is a powerful classifier that is inspired via the biological network of the human spine \cite{kabir2022spinalnet, uysal2021classification}, which is proposed by Kabir et al. \cite{kabir2022spinalnet}. The Spinal structure has achieved state-of-the-art (SOTA) performance in several handwritten digit datasets.

The idea of transfer learning is used to ease the complicated training process of the deep learning method. It used a strong knowledge of the main problem and applied it to a bit differently related topics \cite{albardi2021comprehensive, zhao2022pca, qazani2020adaptive, tajaril2017effects}. The transformation of the knowledge from the previous task improves the performance of the new task. Transfer learning is used in different fields such as entertainment \cite{kuhlmann2007graph}, image processing \cite{dai2008translated}, filtering \cite{li2009transfer},  and etc. Transfer learning by freezing weights of initial layers often directs optimization to a local minimum. Transfer learning without freezing weight is similar to traditional learning \cite{jhong2022expert}. Only the initial weights are transferred. Multiple training with transferred initialization can potentially lead the optimization to the global minima.

Contributions of this paper are as follows:
\begin{enumerate}
     \item Applying transferred initialization for handwritten digit classification.
    \item We apply Spinal fully connected layer with the random initialization, and transfer learned weights of initial layers.
    \item Compared transferred initialization with transfer learning, and results indicate superior performance.
    \item We also discuss transfer learning, data augmentation, and SpinalNet structure.
\end{enumerate}

The rest of the paper is organized as follows. Section II presents several closely related works. Section III presents theoretical details. This section presents the dataset, augmentation, training, SpinalNet, performance parameters. Section IV presents the result. Section V is the concluding section.

\section{Related Works}
There exist numerous works in both deep learning-based COVID-19 diagnosis. COVID-19 was first identified as an outbreak in December 2019. Most of the datasets and competitions related to COVID-19 is published since 2020. Papers in early stages of COVID-19 has received tremendous attentions and citations. Most of the deep learning based COVID-19 diagnosis takes X-ray images as inputs. Diagnosis from X-ray images has reseived areat attention due to convenance. The conventional technique of taking samples from nose for COVID-19 diagnosis seems troublesome to many patients and health personals. 

Several popular deep learning models has proposed for fine-tuned X-ray classification for the diagnosis of coronavirus disease.  Several popular models are Covidx-net \cite{hemdan2020covidx}, Covid-net \cite{wang2020covid},  COVIDiagnosis-Net \cite{ucar2020covidiagnosis},  CoV2-Detect-Net \cite{dixit2021cov2}, SpinalXNet \cite{kumar2021spinalxnet}, Covid-net cxr-s \cite{aboutalebi2021covid}, Corona-net \cite{elbishlawi2021corona}, etc.

Hemdan et al. has proposed Covidx-net \cite{hemdan2020covidx}. Their proposed Covidx-net demonstrated seven different DNN architectures. All of those architectures are previously proposed, and very popular architectures. They applied those DNNs for coronavirus diagnosis.

Wang et al. proposed COVID‑Net \cite{wang2020covid}. They customized a conventional DNN for COVID-19 and the pneumonia dataset. They did a rigorous performance analysis of their proposed DNN. 

Ucar et al. proposed COVIDiagnosis-Net \cite{ucar2020covidiagnosis}. They demonstrated an ML-based novel structure that outperformed existing models of that time. They selected SqueezeNet and tuned for the coronavirus disease diagnosis with the Bayesian optimization. 

Dixit et al. proposed CoV2-Detect-Net \cite{dixit2021cov2}. They have applied feature extraction and K-means clustering for pre-processing data. They also proposed a novel feature optimization technique relying on particle swarm optimization and a hybrid differential evolution algorithm. Finally, a support vector machine classifier distinguishes the samples. They applied their model to publicly available data.

Kumar et al. proposed SpinalXNet \cite{kumar2021spinalxnet}. They applied transfer learning with a modified fully connected layer for classifying X-ray images. They modified the fully connected layed to a SpinalNet layer and achieved superior performance compared to the traditional counterpart.

Aboutalebi et al. proposed Covid-net cxr-s \cite{aboutalebi2021covid}. they have applied DNNS on chest X-ray (CXR) images. Their DNN was applied for predicting the airspace severity of a coronavirus-positive patient.

Elbishlawi proposed CORONA-Net \cite{elbishlawi2021corona} for COVID-19 diagosis. Their proposed DNN was divided into two phases: (1) The reinitialization phase and (2) the classification phase. They also applied their proposed DNN on a publicly available dataset.

As different researchers have applied their proposed DNNs in different datasets, their performance is not comparable. We applied our proposed methodology on the dataset provided by Society for Imaging Informatics in Medicine (SIIM) \cite{vaya2020bimcv}. They launched a competition while they made the dataset publickly available on mid 2021.

\section{Methodology}
The MNIST is one of the most popular datasets in image vision. The dataset contains grey-scale images. However, most of the transfer learning models are trained on RGB data. That might be a reason for the limited conducted studies on transfer learning-based hand-written digit recognition. Therefore, we show the transfer learning process in MNIST on a Kaggle notebook. A deep learning model provides a good result depending on several factors. Many datasets require special augmentation to get an eye-catching result. MNIST dataset requires several augmentations.

\begin{figure}
\begin{center}
\includegraphics[clip, trim=0 0 0 0, width=4.5cm] {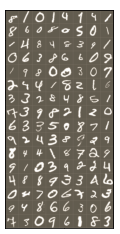}
\caption{\label{MNIST_data} The several samples in MNIST Dataset. When different people write digits, they in a slightly different way. That creates uncertainty in the data.}
\end{center}
\end{figure}

\begin{figure}
\begin{center}
\includegraphics[clip, trim=0 0 0 0, width=7cm] {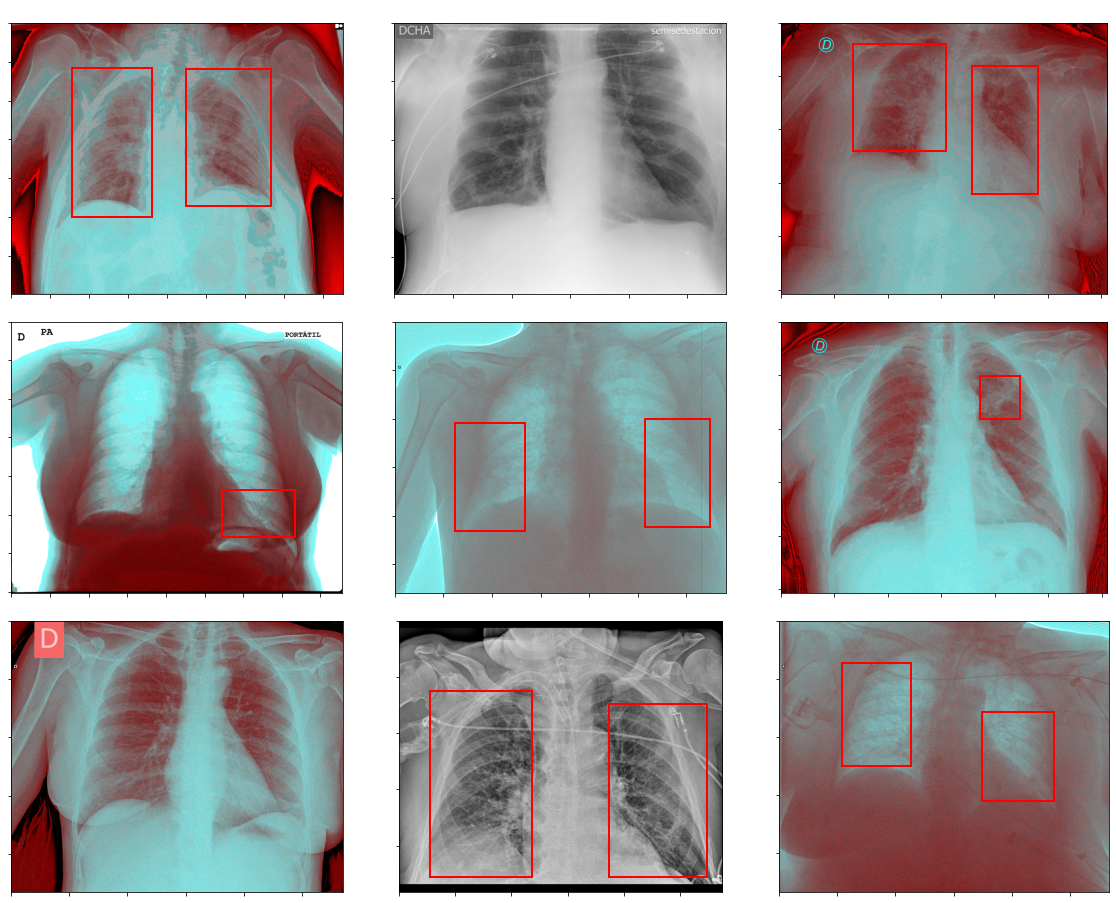}
\caption{\label{COVID_data} The several samples with opaque areas in SIIM-FISABIO-RSNA COVID-19 Detection Dataset. The challenge expects both the detection of opaque images and localization of opaque areas. }
\end{center}
\end{figure}

\subsection{Dataset: MNIST}
The MNIST dataset contains ten thousand test images and sixty thousand training images. Images are in the grey-scale format with 28x28 pixel dimensions. There are ten classes of images starting from 0 to 9. Fig. \ref{MNIST_data} presents the several samples in MINST dataset.

As the original MNIST dataset has training and test sets, we split the training data into training and validation subsets to avoid overfitting. We keep 90\% of the data in the training set, and we move 10\% of the data into the validation set.

\subsection{Dataset: COVID-19 Detection}
The Dataset was provided by Society for Imaging Informatics in Medicine (SIIM) \cite{vaya2020bimcv}. The images were in digital imaging and communications in medicine (DICOM) format. There exist 6334 samples in this dataset. First, we convert DICOM images to RGB images of 512$\times$512 size. Red, Green, and Blue components present magnitude, edge magnitude, and edge angle of DICOM images.

\subsection{Data Augmentation}
Most of the PyTorch transfer learning models are pre-trained on 224x224 sized images \cite{qazani2021time}. However, the MNIST data contains 28$\times$28 sized images. Moreover, further augmentations create fewer noises while the image size is larger. We tested several resizing options in our local computers, and 112$\times$112 is found to be an optimal size for the MNIST dataset. Initially, we also perform random rotation and random perspective to augment the data. The data augmentation varies randomly for different images. As a result, the NN training becomes more robust, and the NN becomes prepared for slightly different test images.

To augment train images of COVID-19 dataset we did a center crop of 470$\times$470 size. We performed random rotations of ten degrees. We also applied random perspective, random horizontal flip, and random greyscale functions of the PyTorch library. Finally, we did a center crop of 448$\times$448 size. We did not augment validation and test images. We resized them to 448$\times$448 size.

\begin{figure}
\begin{center}
\includegraphics[clip, trim=9cm 1cm 9cm 0, width=6cm] {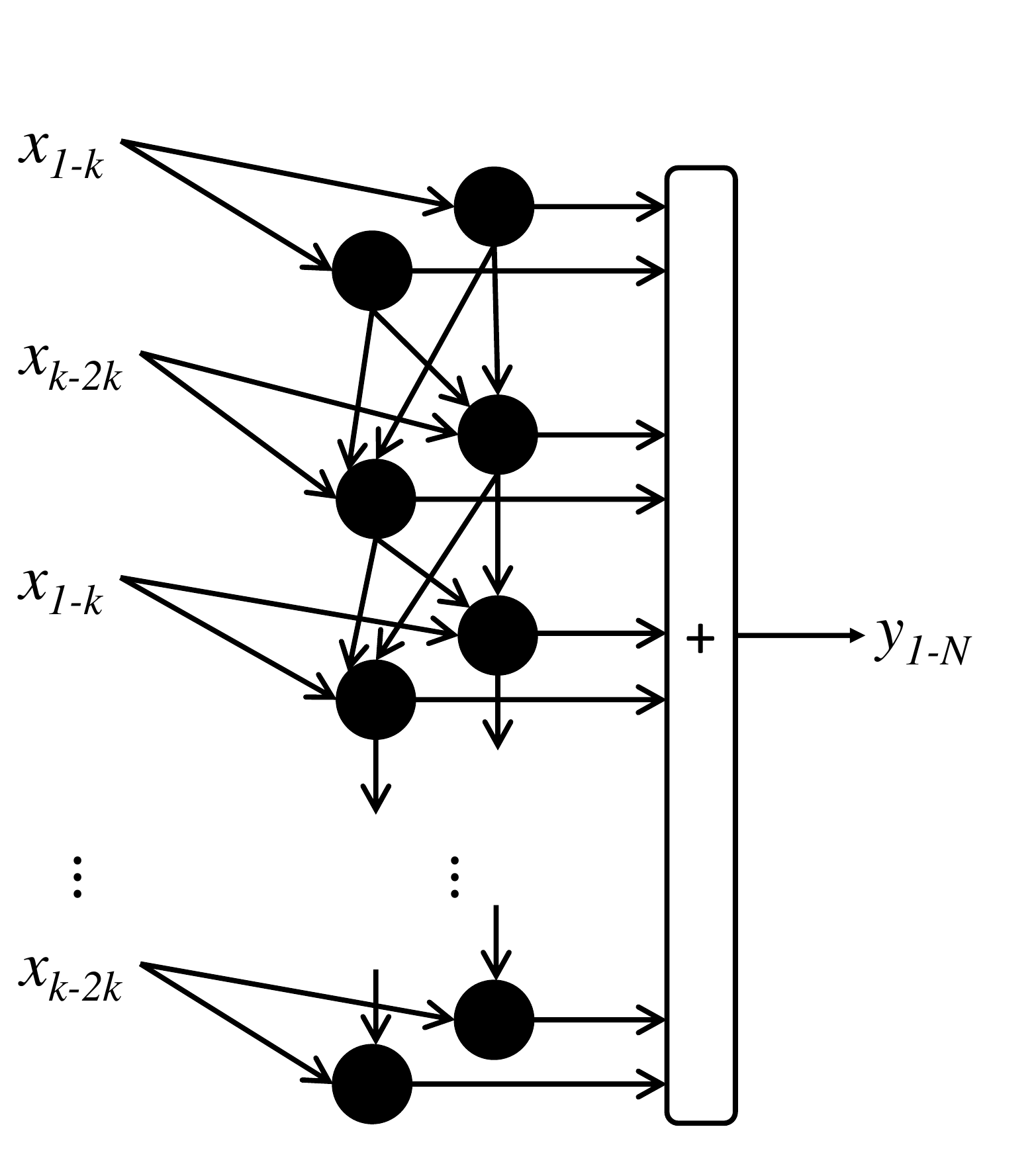}
\caption{\label{Spinal} The structure of SpinalNet. The SpinalNet takes inputs gradually and repetitively. In this paper, we apply the SpinalNet fully connected layer for transfer learning. }
\end{center}
\end{figure}

\subsection{SpinalNet}
The observation of the human spinal cord reached the development of the SpinalNet by Kabir et al. \cite{kabir2022spinalnet}. The SpinalNet is the powerful classifier method that imitates the duty of the human spinal cord. It receives the inputs gradually, similar to the spinal cord. Also, it sends the processed input data to the global output. It tuns the weight parameters of the network during the training process of the network. Fig. \ref{Spinal} shows the schematic structure of the SpinalNet. The network consists of the input rows, multiple hidden layers in the row and the row of outputs. The input of every layer is the input of the previous layer to decrease the number of multiplications without making an underfitting issue. Based on Fig. \ref{Spinal}, the input is divided into two rows to be imported to different hidden layers, respectively.

\subsection{Transfer Learning}
In real life, the human brain gathers experiences through education and many events. Later humans use the learned skills and strategies for accomplishing a new task \cite{kabir2022spinalnet} by seeing a few examples on that task. Then, it is worth mentioning that the transfer is an integral part of the learning process. In this study, the transfer learning technique is applied for training the network to recognize the hand-writing. There are two main issues with applying this method. Initially, the training process of the hand-writing using deep learning techniques needs a massive amount of data which is not always available for the home user researchers. Secondly, the computational load of the training process is very expensive and time-consuming. While using transfer learning, the computational load and training time decreases. Then, the model can be run in a shared server in a short period.

\begin{figure*}
\begin{center}
\includegraphics[clip, trim=0 0 0 0, width=14cm] {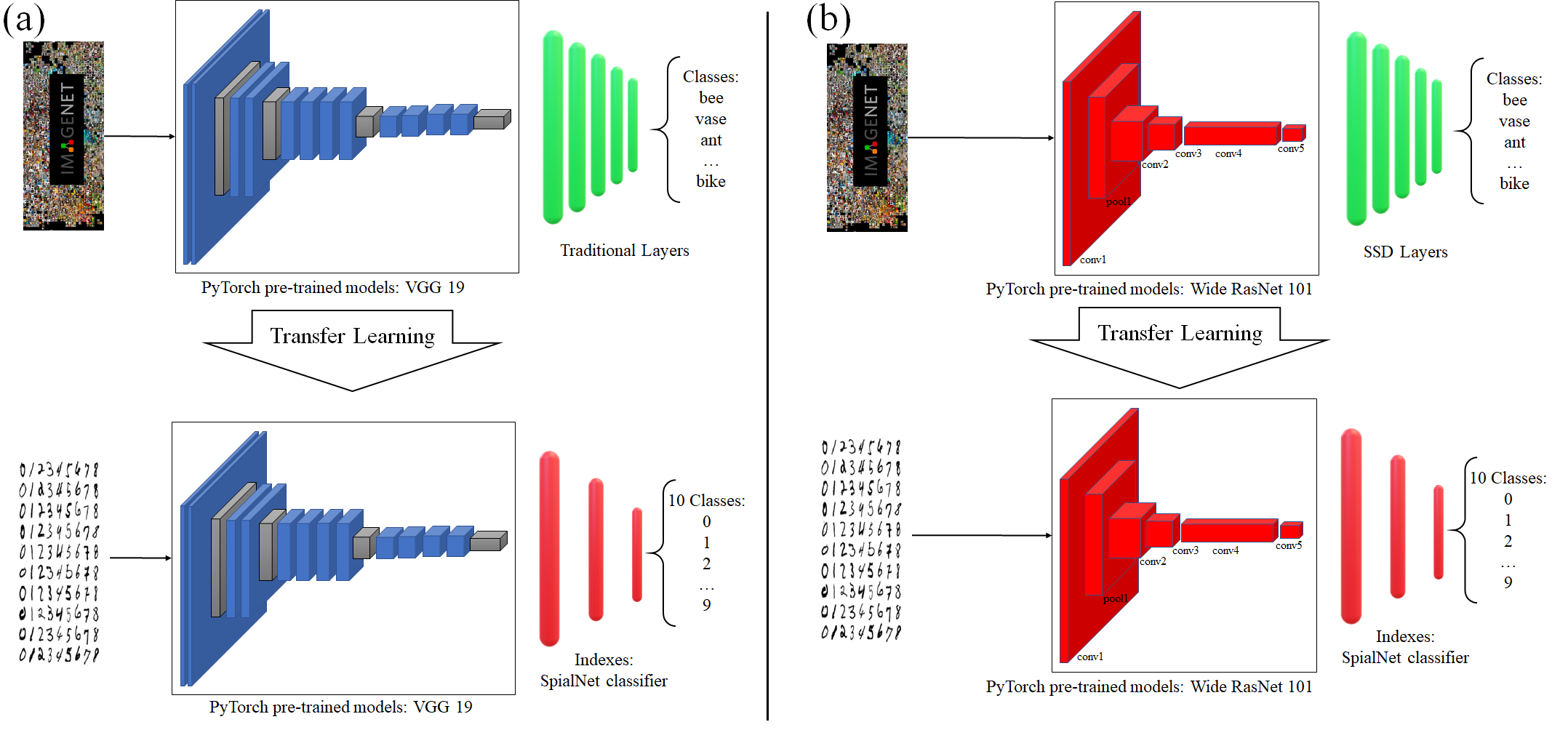}
\caption{\label{TL} The transfer method handwriting detection using based models including (a):VGG-19 ; (b) WideResNet-101. }
\end{center}
\end{figure*}

The model is designed and developed using VGG19\_bn and WideResNet-101 as the base models \cite{kabir2022spinalnet}, which are two of the best models in the ImageNet dataset for classifying the data. Fig. \ref{TL}a-b shows the structure of the proposed transfer learning-based method hand-writing detection using based models including VGG19 and WideResNet-101, respectively. There is a similarity between classifying the image data and hand-writing recognition. It should be noted that the employed base model has been trained using a massive amount of data with many layers and millions of tenable indexes. These indexes and layers are kept constant during the training process of the proposed method. The front line of the proposed transfer learning-based model is fed with the hand-writing images of the number from 0-9. The SpinalNet is employed in the last layer of a VGG network, which resulted in less tenable parameters to classify the hand-writing characters, and this is explained in the next subsection.

\subsection{Performance Matrices}
Popular performance matrices are overall accuracy, cross-entropy loss, confusion matrix, precision, recall, and F1 score. The overall accuracy of test data is the ratio between correctly predicted samples and the total number of samples\cite{chalak2020performance, qazani2020new, qazani2021adaptive}. A confusion matrix is a graphical view that represents which classes are frequently wronged by the model. Also, the confusion matrix shows the wrongly predicted class.    

The precision of Class-X is the ratio between the number of samples the model predicts correctly as Class-X and the total number of samples the model predicts as Class-X. The recall of Class-X is the ratio between the number of samples the model predicts correctly as Class-X and the total number of samples labeled as Class-X. F1 score is the harmonic mean of recall and precision\cite{qazani2020optimising, qazani2021optimal}.

\section{Results and Discussions}
In this section, the proposed model in Section II is verified and validated in simulation environment, and the results are provided to prove the significant of the proposed method. We download images and pre-trained models with the help of the PyTorch command. Also, we apply a popular PyTorch transfer learning code to train the network.  In the code, we first write data loader functions with relevant transformations. We observe augmented images before the training. We declare SpinalNet with a layer width of 1024 while using VGG19\_bn. We declare SpinalNet with a layer width of 20 while using WideResNet-101. The number output is equal to the number of classes. The MNIST dataset has ten classes.

We train the NN in two stages. The initial training has a learning rate of 0.01. We apply ‘SGD’ as an optimizer, ‘lr\_schedular’ as the schedular, and ‘CrossEntropyLoss’ criterion. The momentum of SGD is set to 0.9. Step size and the multiplicative factor ‘gamma’ are set to 7 and 0.1, respectively. The initial training takes ten epochs. The final training is training with a lower learning rate (0.001). Values of other parameters remain the same as the values obtained from initial training. It should be noted that the proposed methods using the transfer learning technique are coded and implemented in the Kaggle notebook, which is available to be reviewed in notebook\footnote{https://www.kaggle.com/dipuk0506/transfer-learning-on-mnist}.

Table \ref{TAB_1} and Table \ref{TAB_2} present the results using VGG19\_bn and WideResNet-101 as a base model with the implementation of traditional and spinal end layers. The average accuracy of the proposed method using VGG19\_bn improved 0.17\% and 0.04\% as compared with those of WideResNet-101 using traditional and spinal end layers as classifiers, respectively. Also, the top accuracy of the proposed method using VGG19\_bn improved 0.12\% and 0.04\% as compared with those of WideResNet-101 using traditional and spinal end layers as classifiers, respectively. It proves the efficiency of the VGG19\_bn compared with WideResNet-101 as a base model using the transfer learning technique. Moreover, SpinalNet is able to increase the average accuracy of the model by 0.01\% and 0.14\% compared with the traditional end layer using VGG19\_bn and WideResNet-101, respectively. Also, the top accuracy value increases 0.01\% and 0.09\% using SpinalNet as compared with those of the traditional end layer using VGG19\_bn and WideResNet-101 as a base model, respectively.

Transfer learning(TL) keeps good accuracy on average but transfer initialization (TI) brings better top accuracy. If someone tries about 5 times with both TL and TI, a model with TI will likely give the best accuracy. TL may bring the optimization to a local minimum where TI has a chance of getting the global minima. We achieved both higher top accuracy and average accuracy for the COVID-19 dataset using TI.

\begin{figure}
\begin{center}
\includegraphics[clip, trim=0 0 0 0, width=8.5cm] {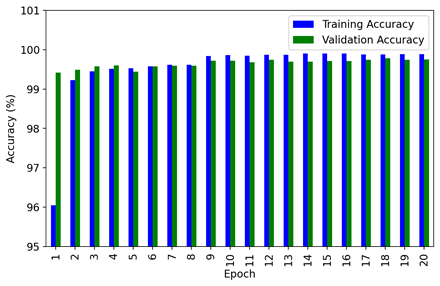}
\caption{\label{Epoch} The accuracy vs epoch plot for WideResNet-101 (Spinal FC) on MNIST dataset. }
\end{center}
\end{figure}

Fig. \ref{Epoch} shows the training and validation accuracy of the proposed transfer learning method on the MNIST dataset, including WideResNet-101 as a base model and SpinalNet as a classifier for 20 epochs. The training accuracy at the first epoch was 96\% which is increased to reach about 99.8\% at the twentieth epoch. Also, the validation accuracy of the WideResNet-101 base model using the SpinalNet classifier was 99.4\% at the first epoch, which increases to 99.71\% at the twentieth epoch.

\begin{figure}
\begin{center}
\includegraphics[clip, trim=0 0 0 0, width=8.5cm] {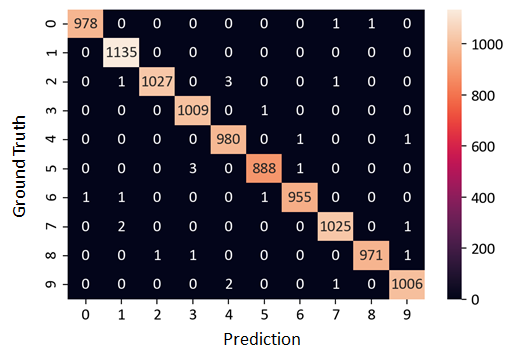}
\caption{\label{Cmat}The sample confusion matrix of the prediction on the MNIST test dataset with VGG19\_bn Spinal FC (Accuracy=99.74\%). }
\end{center}
\end{figure}

Confusion matrix in multiclass classification helps future researchers in understanding the uncertainties in data and models. One AI model may provide good overall accuracy, but the user of the model may not know that where the model might fail. Also, while people are writing digits, they have vibrations in their hands. One person may write f (4), and another person may recognize it as (9). There exist uncertainty and variance in human brains too. Seeing the confusion matrix of a NN in a dataset, one can understand the strength and weaknesses of the models in different input domains. Seeing the confusion matrix of different NN on the dataset, one can understand closely related classes where NNs may fail. Fig. \ref{Cmat} illustrates the sample confusion matrix of the prediction with VGG19\_bn Spinal FC. Based on the presented results, it is obvious that the accuracy of the presented network is 99.74\%. It was able to predict the 9974 true results out of ten thousand test samples. In other words, the error rate of the mentioned network is 0.26\%. Fig. \ref{Cmat} reveals the sensitivity of the VGG19\_bn Spinal FC for different classes. For instance, the recall of the recently developed algorithm for 0, 4, 7, and 9 are 99.80\%, 99.80\%, 99.71\%, and 99.70\%, respectively. In addition, the precision of the different classes such as 0, 4, 7, and 9 are 99.90\%, 99.49\%, 99.71\%, and 99.70\%, respectively. It should be noted that precision is used to show the true prediction rate of the network for different classes. Lastly, the prevalence of the different classes can be calculated based on the presented results in Fig. \ref{Cmat}. It shows the occurrence chance of each class in every trial.

\begin{table} 
		\caption{Accuracies of trained DNNs on MNIST Dataset with Transfer Learning (TL) and Transferred Initialization (TI).}
		\label{TAB_1}
		\centering
			\begin{tabular}{|c|c|c|c|c|c|c|}	\hline
				 &    & \multicolumn{4}{c|}{Accuracy (\%)} \\  \cline{3-6}
				
				 Method &  Models  & \multicolumn{2}{c|}{Traditional FC} &  \multicolumn{2}{c|}{SpinalNet FC} \\  \cline{3-6}  
				     	& & Average& Top    & Average &  Top \\ \hline
				     	TL&VGG19\_bn &99.66 &	99.73 &	99.67 &	99.73 \\ \cline{2-6}
				     	&WideResNet-101 & 99.57 &	99.60&	99.66&	99.69 \\ \hline
				     	
				     	TI&VGG19\_bn &99.69 &	99.74 &	99.70 &	99.75 \\ \cline{2-6}
				     	&WideResNet-101 & 99.52 &	99.62&	99.66&	99.77 \\ \hline
	    \end{tabular} 
\end{table}

\begin{table} 
		\caption{Accuracies of trained DNNs on SIIM-FISABIO-RSNA COVID-19 Dataset with Transfer Learning (TL) and Transferred Initialization (TI).}
		\label{TAB_2}
		\centering
			\begin{tabular}{|c|c|c|c|c|c|c|}	\hline
				 &    & \multicolumn{4}{c|}{Accuracy (\%)} \\  \cline{3-6}
				
				 Method & Models   & \multicolumn{2}{c|}{Traditional FC} &  \multicolumn{2}{c|}{SpinalNet FC} \\  \cline{3-6}  
				     	& & Average& Top    & Average &  Top \\ \hline
				     	TL&VGG19\_bn &72.15 &	73.29 & 73.68 &	73.92 \\ \cline{2-6}
				     	&WideResNet-101 &74.87 &75.55 & 75.49 &	75.89 \\ \hline
				     	
				     	TI&VGG19\_bn &76.52 &	76.91 &76.23 &	77.74 \\ \cline{2-6}
				     	&WideResNet-101 &77.26 &	78.84 &77.88 &	80.01 \\ \hline
	    \end{tabular} 
\end{table}	

\begin{figure}
\begin{center}
\includegraphics[clip, trim=0 0 0 0, width=6cm] {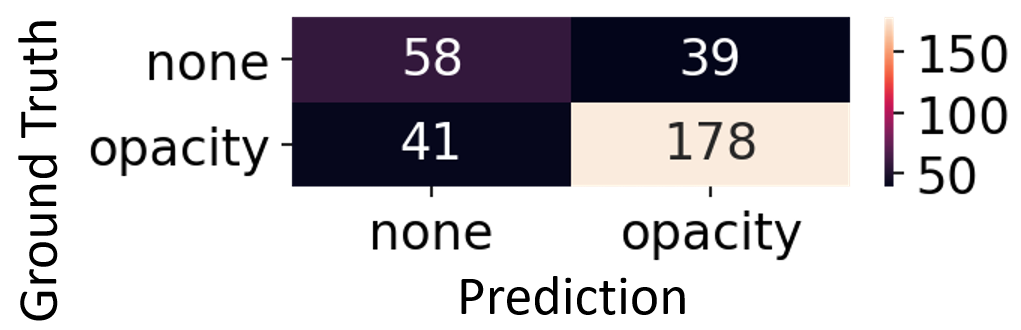}
\caption{\label{CmatCOV}The sample confusion matrix of the prediction on the SIIM-FISABIO-RSNA COVID-19 test dataset with VGG19\_bn Spinal FC (Accuracy=74.68\%). }
\end{center}
\end{figure}

It should be noted that reachable accuracy using the transfer learning method with two base models and two classifiers is reasonable, while the high-tech deep learning methods using complicated structures are able to reach an accuracy of more than 99.79\% \cite{wan2013regularization, kowsari2018rmdl, assiri2020stochastic, mazzia2021efficient, hirata2020ensemble, byerly2020branching}. The developed codes are available on the Kaggle server with execution details$^1$. 

\begin{table} 
		\caption{Precision, Recall, and F1-Scores for Different Classes of the MNIST test dataset with VGG19\_bn Spinal FC. }
		\label{P_recall}
		\centering
			\begin{tabular}{|c|c|c|c|c|c|c|}	\hline
			Class & TP & FP & FN & Precision & Recall&F1 Score \\ \hline 
			0 & 978 & 1 & 2 & 0.9990 & 0.9980 &0.9985\\ \hline 
			1 & 1135 & 4 & 0 & 0.9965 & 1.0 & 0.9982 \\ \hline 
			2 & 1027 & 1 & 5 & 0.9990 & 0.9952 & 0.9971 \\ \hline 
			3 & 1009 & 4 & 1 & 0.9961 & 0.9990 & 0.9975 \\ \hline 
			4 & 980 & 5 & 2 & 0.9949 & 0.9980 & 0.9964 \\ \hline 
			5 & 888 & 2 & 4 & 0.9978 & 0.9955 & 0.9966 \\ \hline 
			6 & 955 & 2 & 3 & 0.9979 & 0.9969 & 0.9974 \\ \hline 
			7 & 1025 & 2 & 3 & 0.9981 & 0.9971 & 0.9976\\ \hline 
			8 & 971 & 1 & 3 & 0.9990 & 0.9969 & 0.9979 \\ \hline 
			9 & 1006 & 3 & 3 & 0.9970 & 0.9970 & 0.9970 \\ \hline 
	
	    \end{tabular} 
\end{table}

\begin{table} 
		\caption{Precision, Recall, and F1-Scores for Different Classes of the COVID-19 test dataset with VGG19\_bn Spinal FC. }
		\label{P_recall2}
		\centering
			\begin{tabular}{|c|c|c|c|c|c|c|}	\hline
			Class & TP & FP & FN & Precision & Recall&F1 Score \\ \hline 
			none & 58 & 41 & 39     & 0.5859 & 0.5979 &0.5918\\ \hline 
			opacity & 178 & 39 & 41 & 0.8203 & 0.8128 & 0.8165 \\ \hline

	    \end{tabular} 
\end{table}	

Table \ref{P_recall} presents the performance of trained NNs for individual classes. Columns in this table represent class, true positive (TP), false positive (FP), false negative (FN), precision, recall, and F1 score values. The model performs well for classes 0, 1. and 8. The model performs comparatively poorly for class 4 and class 5. When people write 0, they write clearly. When people write 4, they write in different ways often the model gets confused between 4, 5, 9, and 3, while recognizing handwritten digits.

\section{Conclusion}
The purpose of this research is to present the novel transferred initialization code on the MNIST and COVID-19 datasets. Most transfer learning models in Torch-vision are designed for RGB images. However, they are easily implementable in the grey-scale MNIST data. The existing hand-writing proposed deep learning models are very expensive from a point of computational perspective. It is not possible to develop a highly accurate model (more than 99.60\% accuracy) without using complicated networks with huge computational time during the training process. In this research, the PyTorch pre-trained models (VGG19\_bn and WideResNet-101) are applied in the MNIST dataset for the first time to maintain accuracy while reducing the computational load to recognize the handwriting. We split the training set into train and validation subsets to avoid overfitting the models. The proposed method is developed and validated in the Kaggle notebook, and the results prove the accuracy of 99.77\% without causing any high computational time during the training process of the network. We applied that optimized training process for COVID-19 detection.  Moreover, researchers can use the proposed approach and apply it to different datasets aiming to obtain good performance while not causing a high computational load and training process.

\section*{Acknowledgement}
This research was partially funded by the Australian Research Council through Discovery Projects funding scheme (DP190102181).

\bibliographystyle{IEEEtran}
\bibliography{Ref}

\begin{thebibliography}{10}
\providecommand{\url}[1]{#1}
\csname url@samestyle\endcsname
\providecommand{\newblock}{\relax}
\providecommand{\bibinfo}[2]{#2}
\providecommand{\BIBentrySTDinterwordspacing}{\spaceskip=0pt\relax}
\providecommand{\BIBentryALTinterwordstretchfactor}{4}
\providecommand{\BIBentryALTinterwordspacing}{\spaceskip=\fontdimen2\font plus
\BIBentryALTinterwordstretchfactor\fontdimen3\font minus
  \fontdimen4\font\relax}
\providecommand{\BIBforeignlanguage}[2]{{%
\expandafter\ifx\csname l@#1\endcsname\relax
\typeout{** WARNING: IEEEtran.bst: No hyphenation pattern has been}%
\typeout{** loaded for the language `#1'. Using the pattern for}%
\typeout{** the default language instead.}%
\else
\language=\csname l@#1\endcsname
\fi
#2}}
\providecommand{\BIBdecl}{\relax}
\BIBdecl

\bibitem{wang2012end}
T.~Wang, D.~J. Wu, A.~Coates, and A.~Y. Ng, ``End-to-end text recognition with
  convolutional neural networks,'' in \emph{Proceedings of the 21st
  international conference on pattern recognition (ICPR2012)}.\hskip 1em plus
  0.5em minus 0.4em\relax IEEE, 2012, pp. 3304--3308.

\bibitem{jospin2022hands}
L.~V. Jospin, H.~Laga, F.~Boussaid, W.~Buntine, and M.~Bennamoun, ``Hands-on
  bayesian neural networks—a tutorial for deep learning users,'' \emph{IEEE
  Computational Intelligence Magazine}, vol.~17, no.~2, pp. 29--48, 2022.

\bibitem{qazani2020prepositioning}
M.~R.~C. Qazani, H.~Asadi, S.~Mohamed, and S.~Nahavandi, ``Prepositioning of a
  land vehicle simulation-based motion platform using fuzzy logic and neural
  network,'' \emph{IEEE Transactions on Vehicular Technology}, vol.~69, no.~10,
  pp. 10\,446--10\,456, 2020.

\bibitem{chung2014empirical}
J.~Chung, C.~Gulcehre, K.~Cho, and Y.~Bengio, ``Empirical evaluation of gated
  recurrent neural networks on sequence modeling,'' \emph{arXiv preprint
  arXiv:1412.3555}, 2014.

\bibitem{qazani2019model}
M.~R.~C. Qazani, H.~Asadi, and S.~Nahavandi, ``A model predictive control-based
  motion cueing algorithm with consideration of joints’ limitations for
  hexapod motion platform,'' in \emph{2019 IEEE International Conference on
  Systems, Man and Cybernetics (SMC)}.\hskip 1em plus 0.5em minus 0.4em\relax
  IEEE, 2019, pp. 708--713.

\bibitem{lee2009convolutional}
H.~Lee, R.~Grosse, R.~Ranganath, and A.~Y. Ng, ``Convolutional deep belief
  networks for scalable unsupervised learning of hierarchical
  representations,'' in \emph{Proceedings of the 26th annual international
  conference on machine learning}, 2009, pp. 609--616.

\bibitem{rish2001empirical}
I.~Rish \emph{et~al.}, ``An empirical study of the naive bayes classifier,'' in
  \emph{IJCAI 2001 workshop on empirical methods in artificial intelligence},
  vol.~3, no.~22, 2001, pp. 41--46.

\bibitem{murphy2006naive}
K.~P. Murphy \emph{et~al.}, ``Naive bayes classifiers,'' \emph{University of
  British Columbia}, vol.~18, no.~60, pp. 1--8, 2006.

\bibitem{wang2005support}
L.~Wang, \emph{Support vector machines: theory and applications}.\hskip 1em
  plus 0.5em minus 0.4em\relax Springer Science \& Business Media, 2005, vol.
  177.

\bibitem{pedrammehr2018novel}
S.~Pedrammehr, M.~R.~C. Qazani, and S.~Nahavandi, ``A novel axis symmetric
  parallel mechanism with coaxial actuated arms,'' in \emph{2018 4th
  International Conference on Control, Automation and Robotics (ICCAR)}.\hskip
  1em plus 0.5em minus 0.4em\relax IEEE, 2018, pp. 476--480.

\bibitem{yu2009learning}
C.-N.~J. Yu and T.~Joachims, ``Learning structural svms with latent
  variables,'' in \emph{Proceedings of the 26th annual international conference
  on machine learning}, 2009, pp. 1169--1176.

\bibitem{tong2001support}
S.~Tong and D.~Koller, ``Support vector machine active learning with
  applications to text classification,'' \emph{Journal of machine learning
  research}, vol.~2, no. Nov, pp. 45--66, 2001.

\bibitem{kabir2015bangla}
F.~Kabir, S.~Siddique, M.~R.~A. Kotwal, and M.~N. Huda, ``Bangla text document
  categorization using stochastic gradient descent (sgd) classifier,'' in
  \emph{2015 International Conference on Cognitive Computing and Information
  Processing (CCIP)}.\hskip 1em plus 0.5em minus 0.4em\relax IEEE, 2015, pp.
  1--4.

\bibitem{ciregan2012multi}
D.~Ciregan, U.~Meier, and J.~Schmidhuber, ``Multi-column deep neural networks
  for image classification,'' in \emph{2012 IEEE conference on computer vision
  and pattern recognition}.\hskip 1em plus 0.5em minus 0.4em\relax IEEE, 2012,
  pp. 3642--3649.

\bibitem{jindal2019effective}
I.~Jindal, D.~Pressel, B.~Lester, and M.~Nokleby, ``An effective label noise
  model for dnn text classification,'' \emph{arXiv preprint arXiv:1903.07507},
  2019.

\bibitem{krizhevsky2012imagenet}
A.~Krizhevsky, I.~Sutskever, and G.~E. Hinton, ``Imagenet classification with
  deep convolutional neural networks,'' \emph{Advances in neural information
  processing systems}, vol.~25, pp. 1097--1105, 2012.

\bibitem{hassan2017deep}
A.~Hassan and A.~Mahmood, ``Deep learning for sentence classification,'' in
  \emph{2017 IEEE Long Island Systems, Applications and Technology Conference
  (LISAT)}.\hskip 1em plus 0.5em minus 0.4em\relax IEEE, 2017, pp. 1--5.

\bibitem{mikolov2010recurrent}
T.~Mikolov, M.~Karafi{\'a}t, L.~Burget, J.~Cernock{\`y}, and S.~Khudanpur,
  ``Recurrent neural network based language model.'' in \emph{Interspeech},
  vol.~2, no.~3.\hskip 1em plus 0.5em minus 0.4em\relax Makuhari, 2010, pp.
  1045--1048.

\bibitem{yang2016hierarchical}
Z.~Yang, D.~Yang, C.~Dyer, X.~He, A.~Smola, and E.~Hovy, ``Hierarchical
  attention networks for document classification,'' in \emph{Proceedings of the
  2016 conference of the North American chapter of the association for
  computational linguistics: human language technologies}, 2016, pp.
  1480--1489.

\bibitem{kabir2022spinalnet}
H.~D. Kabir, M.~Abdar, A.~Khosravi, S.~M.~J. Jalali, A.~F. Atiya, S.~Nahavandi,
  and D.~Srinivasan, ``Spinalnet: Deep neural network with gradual input,''
  \emph{IEEE Transactions on Artificial Intelligence}, 2022.

\bibitem{uysal2021classification}
F.~Uysal, F.~Hardala{\c{c}}, O.~Peker, T.~Tolunay, and N.~Tokg{\"o}z,
  ``Classification of shoulder x-ray images with deep learning ensemble
  models,'' \emph{Applied Sciences}, vol.~11, no.~6, p. 2723, 2021.

\bibitem{albardi2021comprehensive}
F.~Albardi, H.~D. Kabir, M.~M.~I. Bhuiyan, P.~M. Kebria, A.~Khosravi, and
  S.~Nahavandi, ``A comprehensive study on torchvision pre-trained models for
  fine-grained inter-species classification,'' in \emph{2021 IEEE International
  Conference on Systems, Man, and Cybernetics (SMC)}.\hskip 1em plus 0.5em
  minus 0.4em\relax IEEE, 2021, pp. 2767--2774.

\bibitem{zhao2022pca}
B.~Zhao, X.~Dong, Y.~Guo, X.~Jia, and Y.~Huang, ``Pca dimensionality reduction
  method for image classification,'' \emph{Neural Processing Letters}, vol.~54,
  no.~1, pp. 347--368, 2022.

\bibitem{qazani2020adaptive}
M.~R.~C. Qazani, H.~Asadi, T.~Bellmann, S.~Mohamed, C.~P. Lim, and
  S.~Nahavandi, ``Adaptive washout filter based on fuzzy logic for a motion
  simulation platform with consideration of joints’ limitations,'' \emph{IEEE
  Transactions on Vehicular Technology}, vol.~69, no.~11, pp. 12\,547--12\,558,
  2020.

\bibitem{tajaril2017effects}
M.~J. Tajaril, S.~Pedrammehr, M.~R.~C. Qazani, and M.~J. Nategh, ``The effects
  of joint clearance on the kinematic error of the hexapod tables,'' in
  \emph{2017 5th RSI International Conference on Robotics and Mechatronics
  (ICRoM)}.\hskip 1em plus 0.5em minus 0.4em\relax IEEE, 2017, pp. 39--44.

\bibitem{kuhlmann2007graph}
G.~Kuhlmann and P.~Stone, ``Graph-based domain mapping for transfer learning in
  general games,'' in \emph{European Conference on Machine Learning}.\hskip 1em
  plus 0.5em minus 0.4em\relax Springer, 2007, pp. 188--200.

\bibitem{dai2008translated}
W.~Dai, Y.~Chen, G.-R. Xue, Q.~Yang, and Y.~Yu, ``Translated learning: Transfer
  learning across different feature spaces,'' \emph{Advances in neural
  information processing systems}, vol.~21, pp. 353--360, 2008.

\bibitem{li2009transfer}
B.~Li, Q.~Yang, and X.~Xue, ``Transfer learning for collaborative filtering via
  a rating-matrix generative model,'' in \emph{Proceedings of the 26th annual
  international conference on machine learning}, 2009, pp. 617--624.

\bibitem{jhong2022expert}
S.-Y. Jhong, P.-Y. Yang, and C.-H. Hsia, ``An expert smart scalp inspection
  system using deep learning,'' \emph{Sensors and Materials}, vol.~34, no.~4,
  pp. 1265--1274, 2022.

\bibitem{hemdan2020covidx}
E.~E.-D. Hemdan, M.~A. Shouman, and M.~E. Karar, ``Covidx-net: A framework of
  deep learning classifiers to diagnose covid-19 in x-ray images,'' \emph{arXiv
  preprint arXiv:2003.11055}, 2020.

\bibitem{wang2020covid}
L.~Wang, Z.~Q. Lin, and A.~Wong, ``Covid-net: A tailored deep convolutional
  neural network design for detection of covid-19 cases from chest x-ray
  images,'' \emph{Scientific Reports}, vol.~10, no.~1, pp. 1--12, 2020.

\bibitem{ucar2020covidiagnosis}
F.~Ucar and D.~Korkmaz, ``Covidiagnosis-net: Deep bayes-squeezenet based
  diagnosis of the coronavirus disease 2019 (covid-19) from x-ray images,''
  \emph{Medical hypotheses}, vol. 140, p. 109761, 2020.

\bibitem{dixit2021cov2}
A.~Dixit, A.~Mani, and R.~Bansal, ``Cov2-detect-net: Design of covid-19
  prediction model based on hybrid de-pso with svm using chest x-ray images,''
  \emph{Information sciences}, vol. 571, pp. 676--692, 2021.

\bibitem{kumar2021spinalxnet}
K.~Kumar, S.~Khanam, M.~M.~I. Bhuiyan, M.~R.~C. Qazani, S.~K. Mondal, H.~Asadi,
  H.~D. Kabir, A.~Khorsavi, and S.~Nahavandi, ``Spinalxnet: Transfer learning
  with modified fully connected layer for x-ray image classification,'' in
  \emph{2021 IEEE International Conference on Recent Advances in Systems
  Science and Engineering (RASSE)}.\hskip 1em plus 0.5em minus 0.4em\relax
  IEEE, 2021, pp. 1--7.

\bibitem{aboutalebi2021covid}
H.~Aboutalebi, M.~Pavlova, M.~J. Shafiee, A.~Sabri, A.~Alaref, and A.~Wong,
  ``Covid-net cxr-s: Deep convolutional neural network for severity assessment
  of covid-19 cases from chest x-ray images,'' \emph{Diagnostics}, vol.~12,
  no.~1, p.~25, 2021.

\bibitem{elbishlawi2021corona}
S.~Elbishlawi, M.~H. Abdelpakey, M.~S. Shehata, and M.~M. Mohamed,
  ``Corona-net: Diagnosing covid-19 from x-ray images using re-initialization
  and classification networks,'' \emph{Journal of Imaging}, vol.~7, no.~5,
  p.~81, 2021.

\bibitem{vaya2020bimcv}
M.~d. l.~I. Vay{\'a}, J.~M. Saborit, J.~A. Montell, A.~Pertusa, A.~Bustos,
  M.~Cazorla, J.~Galant, X.~Barber, D.~Orozco-Beltr{\'a}n,
  F.~Garc{\'\i}a-Garc{\'\i}a \emph{et~al.}, ``Bimcv covid-19+: a large
  annotated dataset of rx and ct images from covid-19 patients,'' \emph{arXiv
  preprint arXiv:2006.01174}, 2020.

\bibitem{qazani2021time}
M.~R.~C. Qazani, H.~Asadi, M.~Al-Ashmori, S.~Mohamed, C.~P. Lim, and
  S.~Nahavandi, ``Time series prediction of driving motion scenarios using
  fuzzy neural networks:* motion signal prediction using fnns,'' in \emph{2021
  IEEE International Conference on Mechatronics (ICM)}.\hskip 1em plus 0.5em
  minus 0.4em\relax IEEE, 2021, pp. 1--6.

\bibitem{chalak2020performance}
M.~R. Chalak~Qazani, S.~Pedrammehr, H.~Abdi, and S.~Nahavandi, ``Performance
  evaluation and calibration of gantry-tau parallel mechanism,'' \emph{Iranian
  Journal of Science and Technology, Transactions of Mechanical Engineering},
  vol.~44, no.~4, pp. 1013--1027, 2020.

\bibitem{qazani2020new}
M.~R.~C. Qazani, H.~Asadi, T.~Bellmann, S.~Perdrammehr, S.~Mohamed, and
  S.~Nahavandi, ``A new fuzzy logic based adaptive motion cueing algorithm
  using parallel simulation-based motion platform,'' in \emph{2020 IEEE
  International Conference on Fuzzy Systems (FUZZ-IEEE)}.\hskip 1em plus 0.5em
  minus 0.4em\relax IEEE, 2020, pp. 1--8.

\bibitem{qazani2021adaptive}
M.~R.~C. Qazani, H.~Asadi, M.~Rostami, S.~Mohamed, C.~P. Lim, and S.~Nahavandi,
  ``Adaptive motion cueing algorithm based on fuzzy logic using online
  dexterity and direction monitoring,'' \emph{IEEE Systems Journal}, 2021.

\bibitem{qazani2020optimising}
M.~R.~C. Qazani, S.~M.~J. Jalali, H.~Asadi, and S.~Nahavandi, ``Optimising
  control and prediction horizons of a model predictive control-based motion
  cueing algorithm using butterfly optimization algorithm,'' in \emph{2020 IEEE
  congress on evolutionary computation (CEC)}.\hskip 1em plus 0.5em minus
  0.4em\relax IEEE, 2020, pp. 1--8.

\bibitem{qazani2021optimal}
M.~R.~C. Qazani, H.~Asadi, and S.~Nahavandi, ``An optimal motion cueing
  algorithm using the inverse kinematic solution of the hexapod simulation
  platform,'' \emph{IEEE Transactions on Intelligent Vehicles}, vol.~7, no.~1,
  pp. 73--82, 2021.

\bibitem{wan2013regularization}
L.~Wan, M.~Zeiler, S.~Zhang, Y.~Le~Cun, and R.~Fergus, ``Regularization of
  neural networks using dropconnect,'' in \emph{International conference on
  machine learning}.\hskip 1em plus 0.5em minus 0.4em\relax PMLR, 2013, pp.
  1058--1066.

\bibitem{kowsari2018rmdl}
K.~Kowsari, M.~Heidarysafa, D.~E. Brown, K.~J. Meimandi, and L.~E. Barnes,
  ``Rmdl: Random multimodel deep learning for classification,'' in
  \emph{Proceedings of the 2nd International Conference on Information System
  and Data Mining}, 2018, pp. 19--28.

\bibitem{assiri2020stochastic}
Y.~Assiri, ``Stochastic optimization of plain convolutional neural networks
  with simple methods,'' \emph{arXiv preprint arXiv:2001.08856}, 2020.

\bibitem{mazzia2021efficient}
V.~Mazzia, F.~Salvetti, and M.~Chiaberge, ``Efficient-capsnet: Capsule network
  with self-attention routing,'' \emph{arXiv preprint arXiv:2101.12491}, 2021.

\bibitem{hirata2020ensemble}
D.~Hirata and N.~Takahashi, ``Ensemble learning in cnn augmented with fully
  connected subnetworks,'' \emph{arXiv preprint arXiv:2003.08562}, 2020.

\bibitem{byerly2020branching}
A.~Byerly, T.~Kalganova, and I.~Dear, ``A branching and merging convolutional
  network with homogeneous filter capsules,'' \emph{arXiv preprint
  arXiv:2001.09136}, 2020.

\end{thebibliography}

\end{document}